\documentclass[12pt,preprint]{aastex}

\newcommand\be{\begin{equation}}
\newcommand\ee{\end{equation}}
\newcommand\ba{\begin{eqnarray}}
\newcommand\ea{\end{eqnarray}}

\newcommand\bB{\textbf{\emph{B}}}
\newcommand\bT{\textbf{\emph{T}}}
\newcommand\bG{\textbf{\emph{G}}}
\newcommand\tomega{{\tilde\omega}}
\newcommand\bn{\begin{enumerate}}
\newcommand\en{\end{enumerate}}

\newenvironment{inlinefigure}{
\medskip
\def\@captype{figure}
\noindent\begin{minipage}{0.999\linewidth}\begin{center}}
{\end{center}\end{minipage}\medskip}

\begin{document}

\title{Effects of Magnetic Fields on the Diskoseismic Modes of Accreting Black Holes}
\author{Wen Fu and Dong Lai}
\affil{Department of Astronomy, Cornell University, Ithaca, NY 14853\\
Email: wenfu, dong@astro.cornell.edu}

\begin{abstract}
The origin of the rapid quasi-periodic variabilities observed in a number
of accreting black hole X-ray binaries is not understood. It has been
suggested that these variabilities are associated with diskoseismic oscillation
modes of the black hole accretion disk. In particular, in a disk with no magnetic
field, the so-called g-modes (inertial oscillations) can be self-trapped
at the inner region of the disk due to general relativistic effects. Real accretion
disks, however, are expected to be turbulent and contain appreciable magnetic fields.
We show in this paper that even a weak magnetic field (with the magnetic energy much
less than the thermal energy) can modify or ``destroy'' the self-trapping zone of disk
g-modes, rendering their existence questionable in realistic black hole accretion disks.
The so-called corrugation modes (c-modes) are also strongly affected when the poloidal
field approaches equal-partition. On the other hand, acoustic oscillations (p-modes),
which do not have vertical structure, are not affected qualitatively by the magnetic
field, and therefore may survive in a turbulent, magnetic disk.
\end{abstract}

\keywords{accretion: accretion disks --- black hole physics --- magnetic fields --- X-ray: binaries
--- hydrodynamics --- MHD}

\section{Introduction}

In recent years, quasi-periodic variability has been observed from
a number of Galactic compact X-ray binary systems. Of particular
interest are several accreting black hole (BH) binaries which show
pairs of quasi-periodic oscillations (QPOs) of fixed frequencies
having ratios close to $2:3$ (for example, GRO J1655-40 shows
$f=300,~450$~Hz; see Remillard \& McClintock 2006 for a review).
The origin of these QPOs is not understood. The fixed frequencies
and frequency ratio led to the suggestion that these QPOs involve
certain nonlinear resonant phenomenon in the disk (e.g., coupling
between the radial and vertical epicyclic oscillations of the disk
fluid element; Kluzniak \& Abramowicz 2002), but so far no fluid
dynamical model producing these resonances has been developed (see
Rebusco 2008 and references therein). Alternatively, it has been
suggested that these QPOs may arise from acoustic oscillations in
an accretion torus (Rezzolla et al. 2003; Lee, Abramowicz \&
Kluziniak 2004), with the oscillation frequencies determined by
the (unknown) radial extent of the torus. Perhaps the
theoretically most developed model for QPOs is the relativistic
diskoseismic oscillation model (Kato \& Fukue 1980; Okazaki et
al.~1987; Nowak \& Wagoner 1991; see Wagoner 1999; Kato 2001 for
reviews), in which general relativistic (GR) effect produces
trapped g-mode (also called inertial mode or inertial-gravity
mode) oscillations in the inner region of the disk. Various
theoretical issues related to this model have been studied, such
as the role of corotational wave absorption (Kato 2003; Li,
Goodman \& Narayan 2003; Silbergleit \& Wagoner 2007) and resonant
mode excitations due to global disk deformation (Kato 2008;
Ferreira \& Ogilvie 2008).

The studies of the oscillation modes of disks/torii, such as those
mentioned above, usually assume that the unperturbed flow is laminar
and has no magnetic field. Real accretion disks, on the other hand,
are highly turbulent due to the nonlinear development of
magnetorotational instability (MRI) (see Balbus \& Hawley 1998 for a
review).  The question therefore arises as to how the MRI-driven
turbulence affects the oscillation modes obtained from hydrodynamical
models and to what extent these trapped modes remain ``valid'' in a
realistic situation. Arras, Blaes \& Turner (2006) attempted to
address this issue by carrying out MHD simulations in the shearing-box
geometry. They showed that axisymmetric standing sound waves give rise
to distinct peaks in the temporal power spectrum, while inertial waves
do not. The discrete frequencies obtained by them were due to the
imposed periodic boundary conditions adopted in the simulations, and
not due to any relativistic effect.  Arras et al. suggested that their
result posses a serious problem for QPO models based on g-modes
(inertial waves). Recently, Reynolds \& Miller (2008) reported on the
results of global simulations of BH accretion disks (using
Paczynski-Wiita pseudo-Newtonian potential) and showed that, while
axisymmetric g-mode oscillations manifest in the hydrodynamic disk
with no magnetic field, they disappear in the magnetic disk where MHD
turbulence develops.

In this paper, we study analytically the effects of magnetic
fields on the relativistic diskoseismic modes in accretion disks
around BHs. We consider both poloidal and toroidal fields and use
local analysis of the full MHD equations to examine how the
magnetic field changes the radial wave propagation diagrams for
various modes.  We show that the trapping region of g-modes can be
easily ``destroyed'' even when the disk field strength is such
that the associated Alfv\'{e}n speed is much smaller than the
sound speed. On the other hand, the propagation characteristics of
p-modes (acoustic oscillations) and c-modes are largely unchanged.
Note that since we assume laminar flows for our unperturbed disks,
we do not directly address the effects of turbulence on disk
modes. However, we believe that our work is relevant to this
issue, since magnetic fields naturally arise in a turbulent disk.

We summarize the basic MHD equations in \S 2 and
review the properties of diskoseismic modes important for our analysis
in \S 3. We examine in \S 4 and \S5 the effects of poloidal field and
toroidal field on those modes, respectively, and discuss the
implications of our result in \S 6.

\section{Basic Equations}

We consider a non-self-gravitating accretion disk, satisfying
the usual ideal MHD equations:
\ba
&& {\partial{\rho} \over \partial t}
+\nabla\cdot(\rho \textbf{\emph{v}})=0,\\
&& {\partial{\textbf{\emph{v}}} \over \partial
t}+({\textbf{\emph{v}}}\cdot\nabla){\textbf{\emph{v}}}
=-\frac{1}{\rho}\nabla \Pi-{\nabla \Phi}+\frac{1}{\rho}\bT,\\
&& {{\partial {\bB}} \over \partial t}=\nabla\times({\textbf{\emph{v}}}
\times{\bB}).
\ea
Here $\rho,\,P,\,\textbf{\emph{v}}$ are the fluid density, pressure and velocity,
$\Phi$ is the gravitational potential, and
\begin{equation}
\Pi \equiv P+\frac{B^2}{8\pi},\quad
\bT \equiv \frac{1}{4\pi}(\bB\cdot\nabla)\bB
\end{equation}
are the total pressure and
the magnetic tension, respectively. The magnetic field $\bB$ also
satisfies the equation $\nabla\cdot\bB=0$. We assume that the fluid obeys the barotropic
equation of state $P=P(\rho)$.

We adopt the cylindrical coordinates $(r, \phi, z)$ which are
centered on the central BH and have the $z$-axis in the direction
perpendicular to the disk plane. The unperturbed background flow
is assumed to be axisymmetric with a velocity field
$\textbf{\emph{v}}=r\Omega(r)\hat \phi$, and magnetic field $\bB=
B_\phi(r)\hat \phi+B_z \hat z$, i.e., $B_z$ is constant while
$B_\phi$ has a radial dependance. Force balance in the unperturbed
flow implies \begin{equation} \bG\equiv
\frac{1}{\rho}\nabla\Pi-\frac{1}{\rho}\bT=\Omega^2 r
\hat{r}-\nabla \Phi. \end{equation}

Consider perturbations of the form $e^{im\phi-i\omega t}$. The
linearized fluid equations are \ba && -i\tomega\delta
\rho+\frac{1}{r}\frac{\partial}{\partial r}(\rho r \delta v_r)
+\frac{im\rho}{r}\delta v_{\phi}+\frac{\partial}{\partial z}(\rho \delta v_z)=0,\\
&& -i\tomega\delta v_r-2\Omega\delta v_{\phi}= G_r\frac{\delta
\rho}{\rho}-\frac{1}{\rho}\frac{\partial}{\partial r}\delta \Pi
+\frac{1}{\rho}(\delta \bT)_{r},\label{eq:linear2}\\
&& -i\tomega\delta v_{\phi}+\frac{\kappa^2}{2\Omega}\delta v_r=
-\frac{im}{\rho r}\delta \Pi+\frac{1}{\rho}(\delta \bT)_{\phi},\\
&& -i\tomega\delta v_z=G_z\frac{\delta
\rho}{\rho}-\frac{1}{\rho}\frac{\partial}{\partial z}\delta \Pi
+\frac{1}{\rho}(\delta \bT)_{z},\label{eq:linear4}\\
&& -i\tomega\delta B_r=\left(\frac{imB_{\phi}}{r}
+B_z\frac{\partial}{\partial z}\right)\delta v_r,\\
&& -i\tomega\delta B_{\phi}=-\frac{\partial}{\partial
r}(B_{\phi}\delta v_r)+B_z \frac{\partial}{\partial z}\delta
v_{\phi}-B_{\phi}\frac{\partial}{\partial z}\delta v_z+r\frac{d\Omega}{dr}\delta B_r,\\
&& -i\tomega\delta B_z=-\frac{B_z}{r}\frac{\partial}{\partial
r}(r\delta v_r)-\frac{imB_z}{r}\delta
v_{\phi}+\frac{imB_{\phi}}{r}\delta v_z, \ea
where
\begin{equation} \tomega=\omega-m\Omega \end{equation} is the
comoving wave frequency, and
\begin{equation}
\kappa^2 \equiv \frac{2\Omega}{r}\frac{d(r^2\Omega)}{dr} \end{equation}
is
the radial epicyclic frequency.
$\delta\rho,~\delta\Pi,~\delta{\textbf{\emph{v}}},~\delta{\bf B}$
are Eulerian perturbations, and $\rho,~\bB$ refer to the
unperturbed flow variables. In addition, for barotropic fluid, we
have \begin{equation} \delta \rho=\frac{1}{c_s^2}\delta
P=\frac{1}{c_s^2}(\delta \Pi-\frac{1}{4\pi}\bB\cdot\delta \bB),
\end{equation} where $c_s$ is the sound speed.

To perform local (WKB) analysis, we consider perturbations with
spatial dependence $e^{ik_r r+ik_z z}$. In the leading-order
approximation, we keep only the radial gradient of $\Omega(r)$ and
$B_\phi(r)$ while assuming that the variation scales of all the
other background quantities are much larger than the wavelength of
the perturbation, i.e., $k_r, k_z \gg 1/r$. The linearized MHD
equations then reduce to
\ba && -{i\tomega \over \rho c_s^2}\delta
\Pi + ik_r\delta v_r +ik_{\phi} \delta v_{\phi}+ik_z \delta v_z
+{i\tomega B_{\phi} \over 4\pi \rho c_s^2}\delta
B_{\phi} + {i\tomega B_z \over 4\pi \rho c_s^2}\delta B_z=0,\label{eq:wkb1}\\
&& -{ik_r \over \rho}\delta \Pi +i\tomega \delta v_r +2\Omega
\delta v_{\phi}+ \left({ik_z B_z \over 4\pi \rho}+{ik_{\phi}
B_{\phi} \over 4\pi \rho}\right)\delta B_r - {B_{\phi} \over 2\pi
\rho
r}\delta B_{\phi}=0,\\
&& -{ik_{\phi} \over \rho}\delta \Pi-{\kappa^2 \over
2\Omega}\delta v_r+i\tomega\delta v_{\phi}+
\frac{(q+1)B_{\phi}}{4\pi\rho r}\delta B_r+\left({ik_z B_z \over
4\pi \rho}+ {ik_{\phi}B_{\phi}
\over 4\pi \rho}\right)\delta B_{\phi}=0,\\
&& -{ik_z \over \rho}\delta \Pi +i\tomega\delta v_z+\left({ik_z
B_z \over 4\pi \rho}+{ik_{\phi} B_{\phi} \over
4\pi \rho}\right)\delta B_z=0,\\
&&
(ik_{\phi} B_{\phi} +ik_z B_z)\delta v_r +i\tomega\delta B_r=0,\\
&& ik_r B_{\phi}\delta v_r -ik_z B_z \delta v_{\phi} +ik_z
B_{\phi} \delta v_z-p\Omega\delta
B_r-i\tomega\delta B_{\phi}=0,\\
&& ik_r B_z\delta v_r +ik_{\phi}B_z \delta
v_{\phi}-ik_{\phi}B_{\phi}\delta v_z -i\tomega\delta
B_z=0,\label{eq:wkb7} \ea where $k_\phi \equiv m/r$. We have
assumed $B_\phi\sim r^q$ and $p \equiv d\ln \Omega/d\ln r$. Note
that in deriving eqs.~(16)-(22), we have dropped the terms
proportional to $G_r$ and $G_z$ in eqs.~(7) and (9): since
$G_r=(\Omega^2-\Omega_K^2)r \sim \Omega_K^2 r (H/r)^2$ and $G_z
\sim \Omega_K^2 z$ (where $\Omega_k$ is the Keplerian frequency,
i.e., the angular frequency in the absence of pressure force),
$G_r$ is much smaller than the other terms in eq.~(7) provided
that $k_r r \gg 1+v_{A\phi}^2/c_s^2$, and $G_z$ is also negligible
if we focus on the mid-plane of the disk.

\section{Hydrodynamic Limit: Diskoseismic Modes}

In the absence of magnetic fields, for $k_z, k_r \gg k_{\phi}$, the perturbed MHD equations (16)-(22)
lead to the dispersion relation:
\begin{equation}
(\tomega^2-\kappa^2)(\tomega^2-k_z^2 c_s^2)=k_r^2 c_s^2 \tomega^2.
\label{eq:disp1}\end{equation}
For $k_z=0$ (or $\tomega^2\gg k_z^2c_s^2$), this becomes
$\tomega^2=k_r^2c_s^2+\kappa^2$, the usual dispersion relation for
spiral density wave; for $\tomega^2\ll k_z^2 c_s^2$, this becomes
$\tomega=\pm \kappa k_z/(k_r^2+k_z^2)^{1/2}$, describing inertial oscillations
(e.g., Goodman 1993).

For an accretion disk, with scale height $H\ll r$, the vertical
dependence of the perturbation is not well described by the plane
wave $e^{ik_z z}$ unless $k_zH\gg 1$. Okazaki et al.~(1987) showed
that for a thin disk, the perturbation equations can be
approximately separated in $r$ and $z$ (see also Nowak \& Wagoner
1991, 1992; Ipser 1994). For example, for vertically isothermal disks with constant
scale height $H$, one finds $\delta P(r,z),\delta v_r(r,z),\delta
v_\phi(r,z) \propto H_n(z/H)$, while $\delta v_z(r,z)\propto
H'_n(z/H)$, where $H_n(Z)$ (with $n=0,1,\cdots$) is the Hermite
polynomials and $H_n'(Z)=dH_n(Z)/dZ$. With this separation of
variables, Okazaki et al. (1987) obtained the dispersion relation for a
given $n$: \begin{equation}
(\tomega^2-\kappa^2)(\tomega^2-n\Omega_\perp^2)=k_r^2 c_s^2
\tomega^2, \label{eq:disp2}\end{equation} where $\Omega_\perp$ is the
vertical epicyclic frequency and is related to $H$ by
$H=c_s/\Omega_\perp$. An important property of relativistic
disks around BHs is that $\kappa$ is non-monotonic.
Three types of trapped modes can be identified (see Fig.~1; see
also Wagoner 1999, Kato 2001 and Ortega-Rodriguez et al. 2006 for
reviews):

(i) P-modes. For $n=0$, waves can propagate in the region where $\tomega^2 > \kappa^2$.
These are acoustic waves (modified by disk rotation), and have also been termed inertial-acoustic
modes. If waves can be
reflected at the disk inner radius ($r_{\rm ISCO}$, the inner-most stable circular
orbit), discrete p-modes can
be self-trapped at the inner-most region of the disk (see Fig.~1a,~1c).

(ii) G-modes. For $n\geq 1$, waves can propagate in the region where
$\tomega^2 < \kappa^2 < n\Omega_\perp^2$ or $\tomega^2>\Omega_\perp^2 >\kappa^2$
(note that $\kappa< \Omega_\perp$ in GR). The former specifies
the g-mode propagation zone: self-trapped g-modes can be maintained in the
region where $\kappa$ peaks (for $m=0$: see Fig.~1b) or in the region where
$\Omega-\kappa/m < \omega/m < \Omega+\kappa/m$ (for $m\neq 0$: see Fig.~1d).\footnote
{Note that non-axisymmetric g-modes with $\omega/m < \Omega(r_{\rm ISCO})$
contain corotation resonance in the wave zone, leading to strong damping of the mode
(Kato 2003; Li et al. 2003; Zhang \& Lai 2006). On the other hand, modes with $\Omega(r_{\rm ISCO}) < \omega/m
< {\rm max}(\Omega+\kappa/m)$ do not suffer corotational damping, and are therefore of great
interest.}
Because these discrete, self-trapped modes do not require special boundary conditions
(e.g., wave reflection at $r=r_{\rm ISCO}$), they have been the focus of most studies of
relativistic diskoseismology. Note that although we call these g-modes (following the terminology
of Kato 2001 and Wagoner 1999), they have no relation to gravity waves, which are driven
by buoyancy. Instead, these modes describe inertial oscillations, and have also been termed
inertial modes (or inertial-gravity modes).

\begin{inlinefigure}
\scalebox{1}{\rotatebox{0}{\plotone{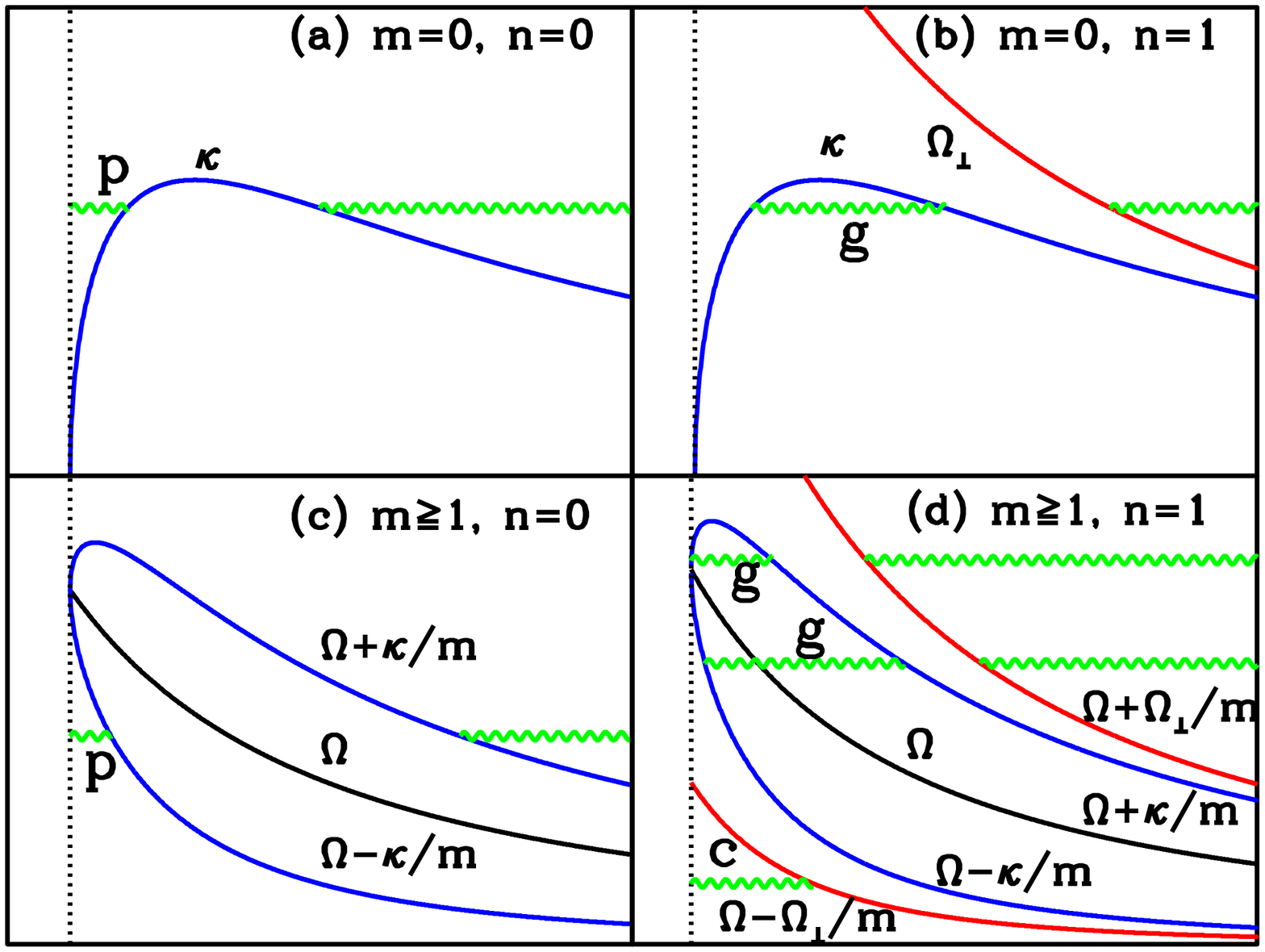}}} \figcaption{Wave
propagation diagram showing various trapped modes in BH accretion
disks: (a) axisymmetric p-mode; (b) axisymmetric g-mode;
(c)non-axisymmetric p-mode; (d) non-axisymmetric g-mode and
c-mode. The curves depict various critical frequencies ($\kappa$,
$\sqrt{n}\Omega_\perp$, $\Omega$, $\Omega \pm \kappa/m$, $\Omega
\pm \sqrt{n}\Omega_\perp/m$), the vertical dotted lines denote the
inner-most stable circular orbit (ISCO). The curvy horizontal
lines specify wave propagation zones and the height of the line is
$\omega$ (for the top panels) or $\omega/m$ (for the bottom
panels) of the mode.}
\end{inlinefigure}

(iii) C-modes. For $n\geq 1$ and $m\geq 1$, the wave propagation condition $\tomega^2 >n\Omega_\perp^2 >\kappa^2$
leads to an additional wave trapping region, where $\omega/m < \Omega-\sqrt{n}\Omega_\perp/m$ (see Fig.~1d).
Note that for spinning BHs, $\Omega_\perp <\Omega$. Clearly, these modes exist only when
$\Omega-\sqrt{n}\Omega_\perp/m >0$ and wave reflection occurs at $r=r_{\rm ISCO}$. Following the previous
works (e.g., Kato 1990 and Silbergleit et al. 2001, who focused on the ``fundamental''
$n=m=1$ mode, corresponding to the Lense-Thirring precession of the inner disk), we call these (``corrugation'') c-modes.

Comparing eqs.~(23) and (24), we see that we can
obtain the radial dispersion relation of different modes by adopting the
vertical ``quantization'' condition $k_z=\sqrt{n}/H$ in
eq.~(23), with $k_z=0$ specifying p-modes. In a generic disk
(e.g., when the disk is not isothermal vertically),
the same ``quantization'' condition would not hold, but we still expect
$k_z\sim 1/H\sim \Omega_\perp/c_s$ for the (vertically) lowest-order g-mode or c-mode.
In the next sections, we will adopt $k_z=\sqrt{\eta}\Omega_\perp/c_s$, with $\eta$ of order
unity, when we study how magnetic fields modify low-order g-modes and c-modes.

Our approach in this paper is based on Newtonian theory. GR effect can be
incorporated into our analysis by using the Paczynski-Witta pseudo-Newtonian potential,
 $\Phi=-M/(r-2M)$. Alternatively, we could simply replace the Newtonian $\Omega,\,\Omega_\perp,\,\kappa$
by their exact general relativistic counterparts (e.g., Okazaki et al. 1987):
\ba
&&
\Omega={{\sqrt {M/r^3}} \over {1+ a\sqrt {M/r^3}}},\\
&&
\Omega_{\perp}=\Omega\left[1 - {{4a M^{1/2}} \over r^{3/2}}+{{3a^2} \over r^2}\right]^{1/2},\\
&&
\kappa=\left[{{M(r^2-6Mr + 8aM^{1/2}r^{1/2}-3a^2)} \over {r^2(r^{3/2} + aM^{1/2})^2}}\right]^{1/2}
\ea
(in geometric units such that $G=c=1$),
where $a = J_s/M$ is the spin parameter of the black hole. In general, $\Omega \geq \Omega_\perp > \kappa$.
In the case of a Schwarzschild BH, $\Omega=\Omega_\perp > \kappa,$ with $\kappa$ peaks at $r=8{\rm M}$
and becomes zero at $r_{\rm ISCO}=6{\rm M}$.
This non-monotonic behavior of the radial epicyclic frequency is preserved for Kerr BHs, and,
as discussed above, is the key ingredient for the existence of trapped diskoseismic modes.

\section{Effect of Poloidal Fields}
We first consider the case of a pure poloidal field, with $B_{\phi}=0$. For $k_z, k_r \gg k_{\phi}$, equations~(16)-(22)
then lead the dispersion relation:
\ba
&&\tomega^6-[(k_z^2+k_r^2)(c_s^2+v_{Az}^2)+k_z^2v_{Az}^2+\kappa^2]\tomega^4\nonumber\\
&&+\left\{k_z^2 v_{Az}^2\left[(k_z^2+k_r^2)(2c_s^2+v_{Az}^2)
+\frac{d\Omega^2}{d\ln r}\right]+\kappa^2k_z^2c_s^2\right\}\tomega^2\nonumber\\
&&-k_z^4v_{Az}^2c_s^2\left[(k_z^2+k_r^2)v_{Az}^2+\frac{d\Omega^2}{d\ln r}\right]=0,\label{eq:mri}
\ea
where $v_{Az}\equiv B_z/\sqrt{4\pi \rho}$. In the incompressible limit, this reduces to the dispersion relation found in,
e.g., Balbus \& Hawley (1991).
For a given ${\bf k}=(k_r, k_{\phi}, k_z)$, equation~(28) admits three branches, corresponding to fast,
slow magnetosonic waves and Alfv\'{e}n wave, all modified by differential rotation. For $k_z \gg k_r$, the Alfv\'{e}n
branch can become unstable when $k_z^2 v_{Az}^2 < -{d\Omega^2}/{d\ln r}.$ This is the well-known MRI (e.g., Balbus \&
Hawley 1998).

\subsection{P-modes}
If $k_z=0$, equation (28) reduces to
\begin{equation}
\tomega^2=\kappa^2+k_r^2(c_s^2+v_{Az}^2).
\end{equation}
This is almost the same expression as in pure hydrodynamic case ($\tomega^2=\kappa^2+k_r^2 c_s^2$), the only
difference being that the sound speed is replaced by fast magnetosonic wave speed, $\sqrt{c_s^2+v_{Az}^2}$.
Thus the basic property of p-modes is not affected by poloidal magnetic fields.

\subsection{G-modes}
For a fixed $k_z=\sqrt{\eta} /H=\sqrt{\eta} \Omega_\perp/c_s$ (see \S3), we can rewrite eq.~(28) as
an expression for $k_r^2$:
\begin{equation}
(c_s^2+v_{Az}^2)k_r^2={{(\tomega^2-\omega_1^2)(\tomega^2-\omega_2^2)(\tomega^2-\omega_5^2)} \over {(\tomega^2-\omega_3^2)(\tomega^2-\omega_4^2)}}.
\end{equation}
The five critical frequencies are given by
\ba
&&\omega_1^2=\eta  (\Omega_{\perp})^2,\\
&&\omega_2^2={1 \over 2}\left[\kappa^2+2\eta(\Omega_{\perp})^2 b^2+\sqrt{\kappa^4+16\eta (\Omega_{\perp}\Omega)^2 b^2}\right]\label{eq:omega2}\\
&&\omega_3^2=\eta (\Omega_{\perp})^2 b^2,\\
&&\omega_4^2=\eta (\Omega_{\perp})^2 {b^2 \over {1+b^2}},\\
&&\omega_5^2={1 \over 2}\left[\kappa^2+2\eta (\Omega_{\perp})^2 b^2-\sqrt{\kappa^4+16\eta (\Omega_{\perp}\Omega)^2 b^2}\right]
\ea
where $b\equiv v_{Az}/c_s$.

Equation~(30) allows us to identify various wave propagation regions ($k_r^2 > 0$). We first consider subthermal
fields, with $b<1$. When $b\lesssim 0.4$ (and with $\eta =1$ for the lowest order g-modes),
the five critical frequencies satisfy
$\omega_1^2 > \omega_2^2 > \omega_3^2 > \omega_4^2 > 0 > \omega_5^2$ in the inner region of the disk.
Thus there are three wave propagation regions:

\ba
&&{\rm Region ~I}:~~ \tomega^2 > \omega_1^2,\\
&&{\rm Region ~II}:  \omega_3^2 < \tomega^2 < \omega_2^2,\\
&&{\rm Region ~III}: \tomega^2 < \omega_4^2.
\ea

\begin{inlinefigure}
\scalebox{1}{\rotatebox{0}{\plotone{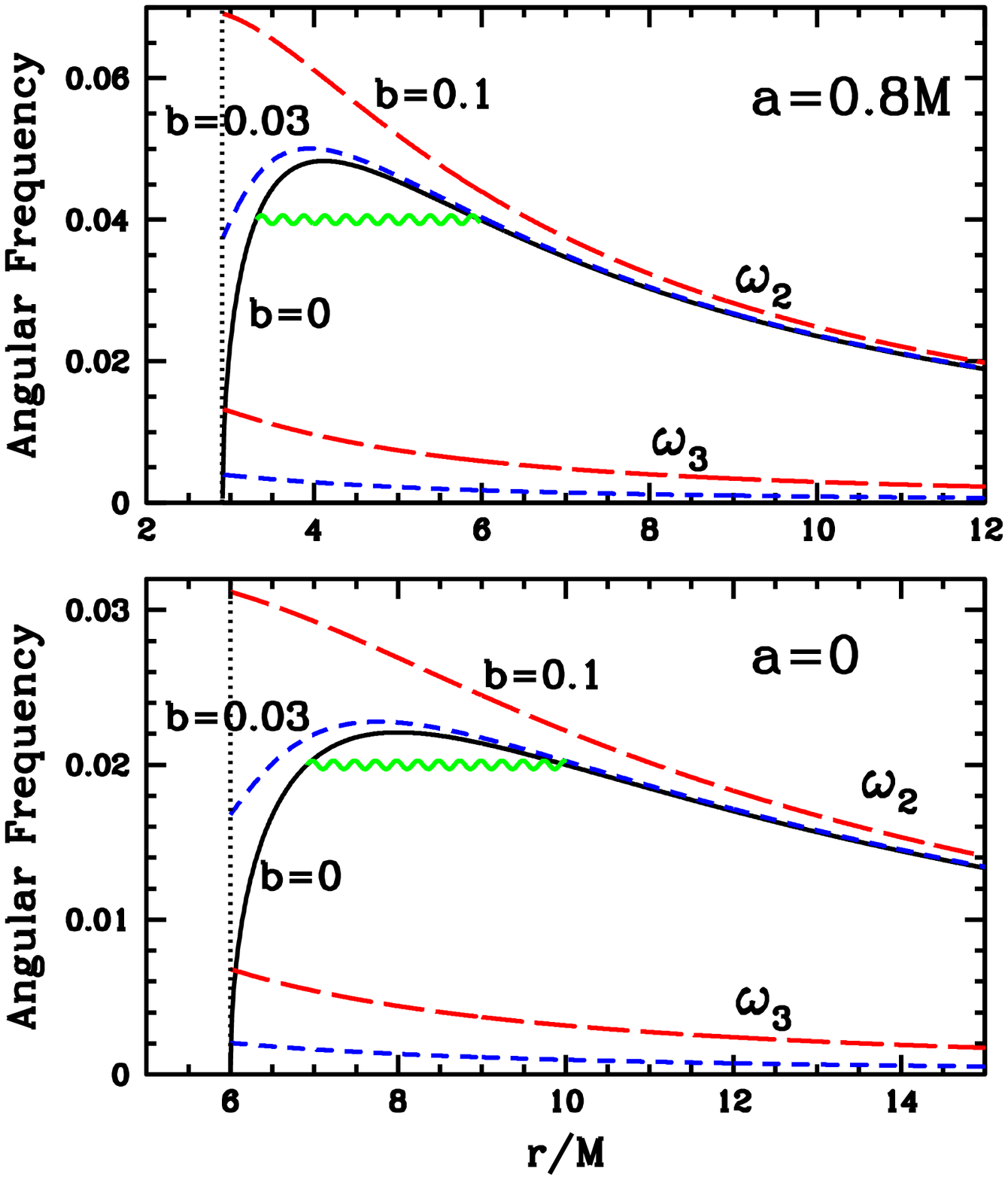}}} \figcaption{The
effect of poloidal magnetic field on the g-mode propagation zone
for $m=0, \eta =1$. In each panel, the upper three curves are
$\omega_2$ (eq.~[32]) and the lower two curves are $\omega_3$
(eq.~[33]). Axisymmetric g-modes of frequency $\omega$ can
propagate in the region where $\omega_3 < \omega < \omega_2$. The
solid line refers to the case of $b=0$, the short-dashed line
$b=0.03$ and the long-dashed line $b=0.1$, where $b\equiv
v_{Az}/c_s$, with $v_{Az}=B_z/\sqrt{4\pi \rho}$ (Alfv\'{e}n speed)
and $c_s$ the sound speed. The vertical dotted lines correspond to
the inner disk radius at ISCO. The curvy horizontal lines specify
the wave propagation zones, and the height of the line is $\omega$
of the mode. The lower and upper panels are for the case of a
Schwarzschild BH ($a=0$) and a Kerr BH ($a=0.8{\rm M}$),
respectively. The angular frequencies are in units of
$M^{-1}=c^3/(GM)$.}
\end{inlinefigure}

Region II corresponds to the original g-mode cavity modified by
the magnetic field;  in the zero field limit, $\omega_3=0$, $\omega_2=\kappa$
and eq.~(37) reduces to $\tomega^2 < \kappa^2$. Fig.~2 depicts the
critical frequencies $\omega_2$ and $\omega_3$ for several values of $b$. This also
serves as the propagation diagram for $m=0$ g-modes (wave can
propagate in region where $\omega_3 < \omega <\omega_2$). We see
that as the magnetic field increases, the g-mode self-trapping zone gradually shrinks
and disappears even when the magnetic field is still very
subthermal (for a Schwarzschild BH, this occurs for $b
\gtrsim 0.08$). More precisely, the g-mode cavity can still exist for large b, but
it now requires a reflection boundary at $r_{\rm ISCO}$.
This behavior can be easily understood by
inspecting eq.~(32):
While $\kappa$ peaks at some radius $r_{\rm max}$, $\Omega_\perp$ and
$\Omega$ both increase monotonically with decreasing $r$. Since
$\Omega_\perp$ and $\Omega$ are much larger than $\kappa$ in the
inner region of the disk, the $2(\Omega_\perp b)^2$ term or
the $4\Omega_\perp \Omega b$ term can dominate over $\kappa^2$
even when $b$ is still small, therefore making the self-trapping zone disappear.
Roughly, this occurs at $b \gtrsim b_{\rm crit} \sim
(\kappa^2/2\Omega_\perp \Omega)_{r_{\rm max}}$.

\begin{inlinefigure}
\scalebox{1}{\rotatebox{0}{\plotone{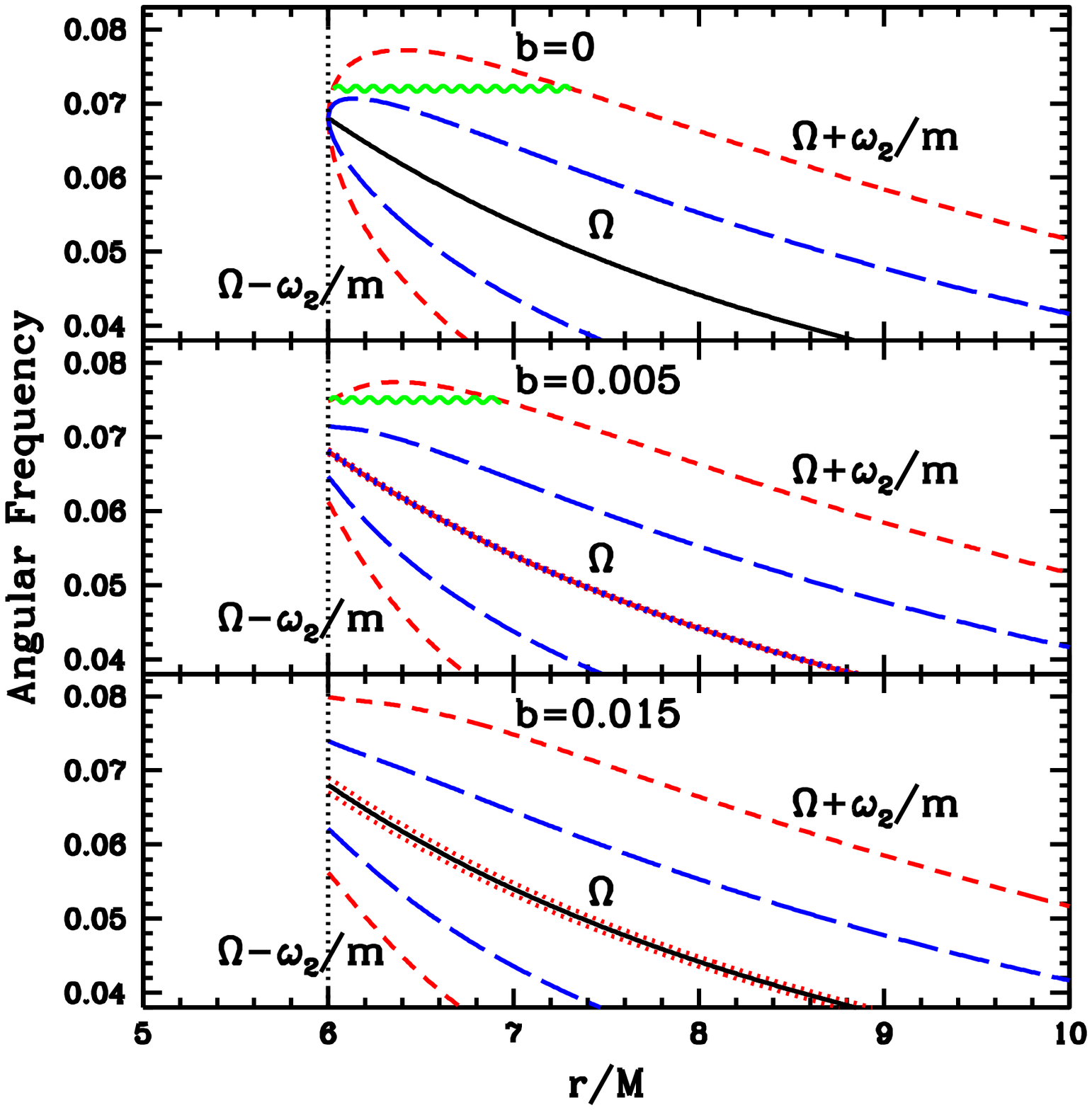}}} \figcaption{The
effect of poloidal magnetic field on the g-mode propagation zone
for $m \neq 0, \eta =1, a=0$. The three panels are for
$b=v_{Az}/c_s =0$, $0.005$ and $0.015$. The solid line,
short-dashed lines and the long-dashed lines show $\Omega$,
$\Omega \pm \omega_2$ and $\Omega \pm \omega_2/2$, respectively.
In the bottom panel, the dotted lines show $\Omega \pm \omega_3$
(In the upper and middle panels, $\Omega \pm \omega_3$ almost
concide with $\Omega$, since $\omega_3=0$ for $b=0$ and $\omega_3
\ll \Omega$ for $b \ll 1$). Non-axisymmetric g-modes can propagate
in the region where $\Omega-\omega_2/m < \omega/m <
\Omega-\omega_3/m$ or $\Omega+\omega_3/m < \omega/m <
\Omega+\omega_2/m$. The vertical dotted lines correspond to the
inner disk radius at ISCO. The curvy horizontal lines in top and
middle panels specify wave propagation zones and the height of the
line is $\omega/m$ of the mode. Note that the self-trapping zone
(depicted in the upper and middle panels) disappears as b
increases. The angular frequencies are in units of
$M^{-1}=c^3/(GM)$.}
\end{inlinefigure}

For non-axisymmetric perturbations ($m \neq 0$), the wave propagation region II is determined by
(i) $\Omega-\omega_2/m < \omega/m < \Omega+\omega_2/m,$ and
(ii) $\omega/m >\Omega+\omega_3/m$ or $\omega/m < \Omega-\omega_3/m$.
Fig.~3 shows the propagation diagram. As mentioned before (see Footnote~1), for $b=0$,
only the modes with $\omega> m\Omega(r_{\rm ISCO})$
are of interest, since otherwise there is a corotation resonance in the wave zone,
leading to strong mode damping (Kato 2003; Li, Narayan \& Goodman 2003; Zhang \& Lai 2006). Thus,
self-trapped g-modes reside around the radius where $\Omega+\omega_2/m$ is the maximum
(and this maximum arises because $\kappa$ depends nonmonotonically on $r$).
We see from Fig.~3 that this g-mode self-trapping region
disappears as $b$ increases. The larger $m$ is, the more fragile is the cavity. For example, the $m=1$ cavity
disappears for $b \gtrsim 0.015$, while for $m=2$, this occurs for $b \gtrsim 0.005$.

\subsection{C-modes}
For $m \neq 0$ and $\eta \sim 1$, equation (30) also describes trapped c-modes. When $b \lesssim 0.4$,
$\omega_1^2$ is the largest among all the critical frequencies and the c-mode propagation zone corresponds to
Region I (see eq.~[36]). Note that since $\omega_1$ is not affected by the magnetic field, the trapping region is
determined by $\omega/m < \Omega-\omega_1/m =\Omega-\Omega_\perp/m$ (for $\eta =1$, see the upper panel of Fig.~4; cf.
Fig.~1d).
When $b \gtrsim 0.4$, the ordering between $\omega_1$ and $\omega_2$ switches and c-modes propagate in the region
where $\tomega^2 > \omega_2^2$, with the trapping zone determined by $\omega/m < \Omega -\omega_2/m$ (see the
bottom panel of Fig.~4). Thus, in the presence of a reflection boundary at $r_{\rm ISCO}$, c-modes
are not affected by the poloidal magnetic field when $b\lesssim 0.4$, but can be appreciably modified
when $b \gtrsim 0.4$.

\begin{inlinefigure}
\scalebox{1}{\rotatebox{0}{\plotone{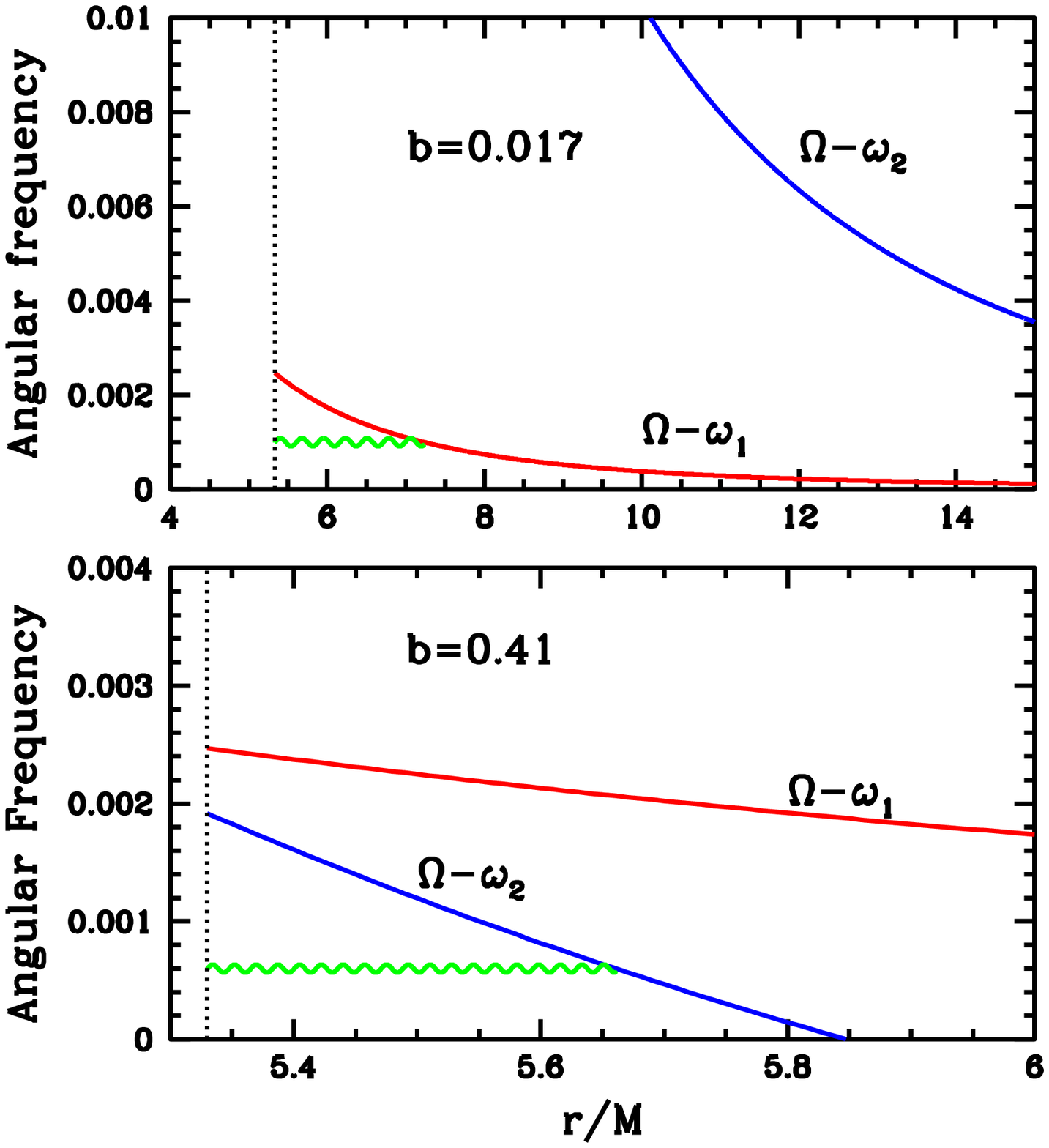}}} \figcaption{The
effect of poloidal magnetic field on c-modes with $m=1$ and
$a=0.2\rm{M}$. The upper panel shows the original c-modes since
$b$ is small and $\omega_1=\Omega_\perp$; in the bottom panel,
with a large $b$, the c-mode trapping zone is instead bounded by
the inner reflection boudary and $\Omega-\omega_2$. The vertical
dotted line refers to the inner disk radius at ISCO. The curvy
horizontal lines specify wave propagation zones and the height of
the line is $\omega$ of the mode.}
\end{inlinefigure}

\begin{inlinefigure}
\scalebox{1}{\rotatebox{0}{\plotone{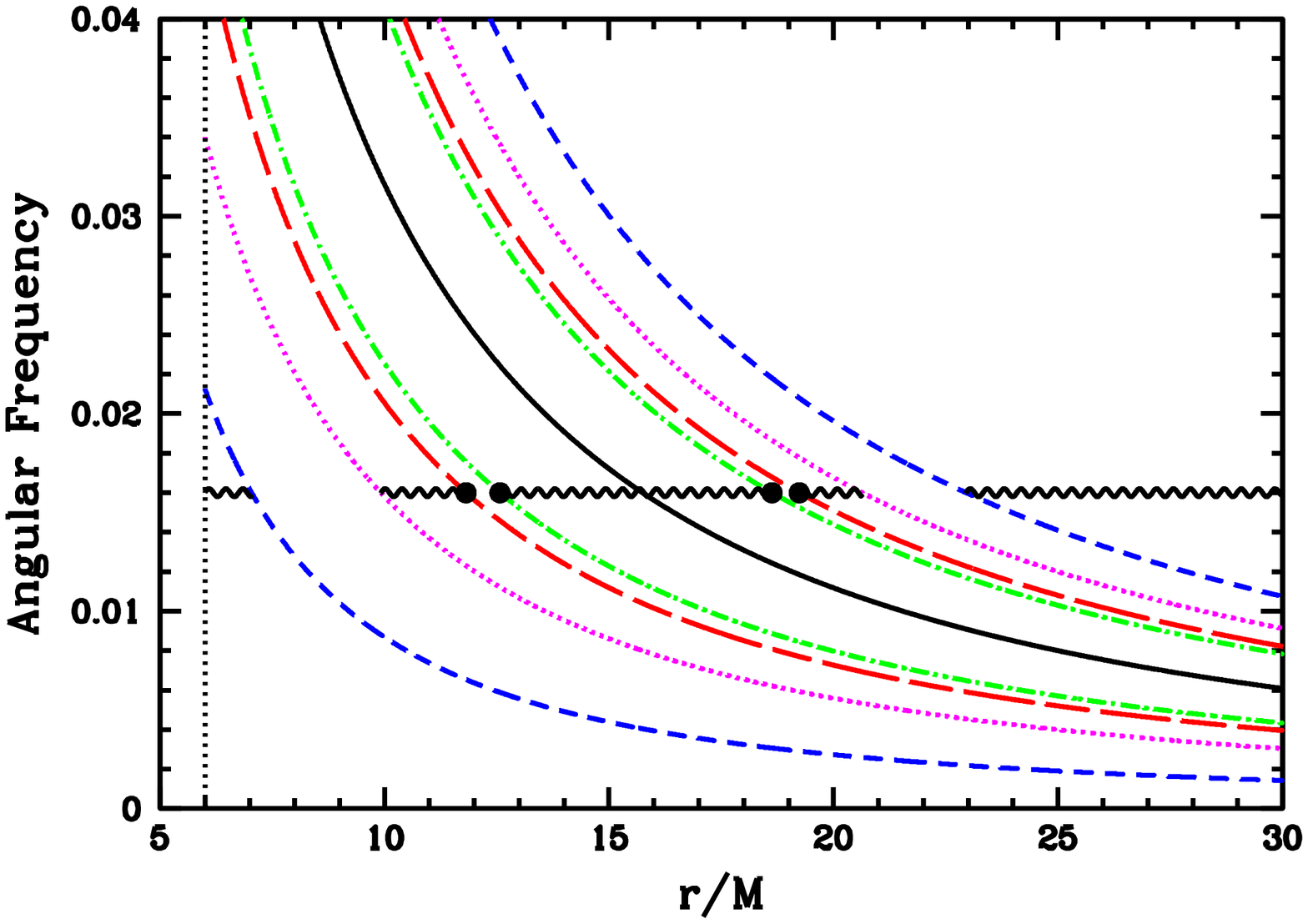}}} \figcaption{Wave
propagation diagram for non-axisymmetric g-modes and c-modes, for
$b=v_{Az}/c_s=0.7$, $m=2$, $\eta =1$, and $a=0$. The solid line,
dot-short dashed lines, long-dashed lines, dotted lines and
short-dashed lines show $\Omega$, $\Omega \pm \omega_4/2$, $\Omega
\pm \omega_3/2$, $\Omega \pm \omega_1/2$ and $\Omega \pm
\omega_2/2$ (eqs.~[31]-[34]), respectively. The vertical dotted
line shows the inner disc radius. The curvy horizontal lines
specify wave propagation zones and the height of the line is
 $\omega/2$ of the mode. The singular points (where $k_r
\rightarrow \infty$) are indicated by filled circles. The angular
frequencies are in units of $M^{-1}=c^3/(GM)$.}
\end{inlinefigure}

From eq.~(30) we can identify other wave propagation zones (see
eq.~[38]). Fig.~5 gives an example, for $m=2$, $b=0.7$. Note that,
except for the c-mode trapping zone discussed above, all the
propagation zones are bounded by at least one ``singular point''
(where $k_r \rightarrow \infty$). Unlike the turning point ($k_r
\rightarrow 0$) associated with wave reflection, wave absorption
is expected to occur at these singular points (see Zhang \& Lai
2006; Tsang \& Lai 2008a and references therein). Thus, the new
wave trapping regions given by eq.~(38) will not lead to
interesting global oscillation modes (Note that in the case of
$b=0.7$, the ordering of five critical frequencies is different
from the one described in \S4.2. However, our conclusion still
holds true, i.e., there is no chance to form a wave trapping zone
bounded by two reflection points other than the c-mode oscillation
region, which is bounded by a reflection point and the ISCO).

\section{Effect of Toroidal Fields}
In this section, we consider the effect of a pure toroidal field,
with $B_z=0$. Various instabilities may exist for such field
geometry, depending upon the rotation profile $\Omega(r)$ and the
magnetic field profile $B_{\phi}(r)$ (e.g., Acheson \& Gibbons 1978; Terquem \&
Papaloizou 1996). Here we focus on how $B_{\phi}$ affects the
diskoseismic modes.

\subsection{P-modes}
With $k_z =0$, equations~(16)-(22) reduces to
\begin{equation}
\tomega^2=\kappa^2+k_r^2(c_s^2+v_{A\phi}^2),
\end{equation}
where $v_{A\phi} \equiv B_{\phi}/\sqrt{4\pi\rho}$.
Thus, the toroidal field affects p-modes in the same way as the poloidal field does (see \S4.1).

\subsection{G-modes}
Since the general dispersion relation for $m \neq 0$ is quite complicated, here we focus on
axisymmetric perturbations.\footnote{Since c-modes necessarily require $m>0$, our analysis here cannot be
applied to c-modes.} With $m=0$, Equations~(16)-(22) lead to
\ba
&&\omega^4-[\kappa^2+(k_z^2+k_r^2)(c_s^2+v_{A\phi}^2)]\omega^2+\kappa^2k_z^2(c_s^2+v_{A\phi}^2)\nonumber\\
&&+2(1-q)v_{A\phi}^2 c_s^2 k_z^2/r^2=0,
\ea
where $q=d\ln B_{\phi}/d\ln r$. Solving for $k_r^2$, we have
\begin{equation}
k_r^2=\frac{(\omega^2-\omega_+^2)(\omega^2-\omega_-^2)}{(c_s^2+v_{A\phi}^2)\omega^2},
\end{equation}
with the two critical frequencies given by
\ba
&&\omega_\pm^2=\frac{\kappa^2+\eta \Omega_\perp^2(1+b_{\phi}^2)}{2}\pm\nonumber\\
&&\frac{1}{2} \sqrt{[\kappa^2-\eta \Omega_\perp^2(1+b_{\phi}^2)]^2
-8(1-q)\eta v_{A\phi}^2 \Omega_\perp^2/r^2},
\ea
where $b_{\phi} \equiv v_{A\phi}/c_s$ and we have used $k_z =\sqrt{\eta} /H=\sqrt{\eta} \Omega_\perp/c_s$
as in \S4. Clearly, for $b_{\phi}=0$, eq.~(42) reduces to eq.~(24).

When $q=1$ (i.e., $B_{\phi} \propto r$), eq.~(42) gives
$\omega_+^2=\eta(\Omega_\perp)^2 (1+b_{\phi}^2),$ and $\omega_-^2=\kappa^2.$
Since $\omega_-^2$ is independent of $B_{\phi}$, the g-mode propagation zone is unaffected no matter
how strong the field is. When $q \neq 1$, as long as $v_{A\phi} \ll \Omega_\perp r$ (which is
valid in most disk situations), the $8(1-q)\eta v_{A\phi}^2\Omega_\perp^2/r^2$ term in eq.~(42) represents only
 a small correction, i.e., $\omega_-^2$ is still very close to $\kappa^2$. Thus for
general toroidal field satisfying $v_{A\phi} \ll \Omega_\perp r$, the axisymmetric
g-mode propagation zone is not affected by the magnetic field.

\section{Summary and Discussion}

In this paper we have studied the effects of both poloidal and
toroidal magnetic fields on the diskoseismic modes in BH accretion
disks. Previous works by Kato, Wagoner and others have been based on
hydrodynamic disks with no magnetic field. The key finding of our
paper is that the g-mode self-trapping zone (which arises from GR
effect) disappears when the disk contains even a small poloidal
magnetic field, corresponding to $v_{Az}/c_s =0.01-0.1$ (see Fig.~2-3;
$v_{Az}$ is the Alfv\'{e}n speed and $c_s$ is the sound speed). It is
well-known that the combination of a weak poloridal field and
differential rotation gives rise to MRI, making real astrophysical
disks turbulent.
Earlier numerical simulations indicated that the magnetic
field grows as MRI develops, until it saturates at $v_{Az}/c_s \sim 0.1-1$,
with the toroidal field stronger than the poloidal field by a
factor of a few (see, e.g., Hawley et al. 1996; Balbus \& Hawley
1998). Recent simulations showed that the turbulent state depends
strongly on the net magnetic flux through the disk (e.g., Fromang
\& Papaloizou 2007; Simon, Hawley \& Beckwith 2008). In any case,
it is likely that the magnetic field in a turbulent disk is large
enough to ``destroy'' the g-mode self-trapping zone.

Thus, the g-mode properties (including the frequencies and
excitations) derived from hydrodynamical models are unlikely to be
applicable to real BH accretion disks. The disappearance of the g-mode
trapping zone might also explain why Arras et al. (2006) and Reynolds
\& Miller (2008) did not see any global g-modes in their MHD
simulations.

As mentioned in \S1, g-mode oscillations have been considered a
promising candidate to explain QPOs in BH X-ray
binaries. Theoretically, these modes are appealing because in
hydrodynamic disks their existence depends on general relativistic
effect and does not require special disk boundary conditions. Our
analytical results presented in this paper, together with recent
numerical simulations (Arras et al. 2006; Reynolds \& Miller 2008),
suggest that magnetic fields and turbulence associated with real
accretion disks can change this picture significantly.

While g-modes can be easily modified or ``destroyed'' by magnetic
fields, our analysis showed that p-modes are not affected
qualitatively. The magnetic field simply changes the sound speed to
the fast magnetosonic wave speed and leaves the p-mode propagation
diagram unchanged. We also showed that a weak poloidal field
($v_{Az}/c_s \ll 1$) does not affect the c-mode propagation zone,
although a stronger field modifies it. Our results therefore suggests
that global p-mode oscillation is robust and may exist in real BH
accretion disks, provided that partial wave reflection at the disk
inner edge can be achieved.\footnote{Kato (2001) has discussed why
such reflection may be possible.} Of particular interest is the
non-axisymmetric p-modes, since they may be excited by instabilites
associated with corotation resonance (Tsang \& Lai 2008a, b).

\begin{acknowledgements}
We thank David Tsang for useful discussion. This work has been supported in part by NASA Grant
NNX07AG81G and by NSF grant AST 0707628.
\end{acknowledgements}


\end{document}